\documentclass[a4paper,11pt,3p,final,twoside]{elsarticle}
\usepackage{amsfonts}
\usepackage[latin1]{inputenc}
\usepackage{times}
\usepackage[T1]{fontenc}
\usepackage{color}
\usepackage{xspace}
 \usepackage[dvips,
 colorlinks,linkcolor=webgreen,filecolor=webbrown,citecolor=webgreen,
 breaklinks=true, urlcolor=webbrown
 ]{hyperref}
\usepackage{amsmath}
\usepackage{amssymb}%
\usepackage{array}
\definecolor{webgreen}{rgb}{0,.5,0}
\definecolor{webbrown}{rgb}{.6,0,0}
\definecolor{webyellow}{rgb}{0.98,0.92,0.73}
\definecolor{webgray}{rgb}{.753,.753,.753}
\definecolor{webblue}{rgb}{0,0,.8}
\graphicspath{{./graphics/}{.}{./figs/}}
\begin{document}

\begin{frontmatter}
\setlength{\hsize}{16.5cm}



\title{Source Coding with Escort Distributions and Rényi Entropy  Bounds}


\author{J.-F. Bercher}
\ead{jf.bercher@esiee.fr}
\address{
Université Paris-Est, LIGM, UMR CNRS 8049, ESIEE-Paris\\
5 bd Descartes, 77454 Marne la Vallée Cedex 2, France}



\begin{abstract}
We discuss the interest of escort distributions and Rényi entropy in the context of source coding.
We first recall a source coding theorem by Campbell relating a generalized measure of length to the Rényi-Tsallis entropy. We show that the associated optimal codes can be  obtained using considerations on escort-distributions. We propose a new family of measure of length involving escort-distributions and we show that these generalized lengths are also bounded below by the Rényi entropy. Furthermore, we obtain that the standard Shannon codes lengths are optimum for the new generalized lengths measures, whatever the entropic index. Finally, we show that there exists in this setting  an interplay between standard and escort distributions. 
\end{abstract}


\end{frontmatter}





\thispagestyle{empty}
\section{Introduction}
Rényi and Tsallis entropies extend the standard Shannon-Boltzmann entropy, enabling to build generalized thermostatistics, that include the standard one as a special case. This has received a very high attention and there is a wide variety of applications where experiments, numerical results and analytical derivations fairly agree with these new formalisms \cite{tsallis_introduction_2009}. These results have also raised interest in the general study of information measures and their applications.  The definition of Tsallis entropy was originally inspired by multifractals whereas the Rényi entropy is an essential ingredient \cite{harte_multifractals:_2001,jizba_world_2004}, e.g. via the definition of the Rényi dimension. 
For a distribution $p$ of a discrete variable with $N$ possible microstates, the Rényi entropy of order $\alpha$, with $\alpha\geq 0$, is defined by
\begin{equation}
 H_\alpha(p)=\frac{1}{1-\alpha}\log \sum_{i=1}^N p_i^\alpha.
\label{eq:renyi}
\end{equation}
By L'Hospital rule, for $\alpha=1$, we recover the Shannon entropy
\begin{equation}
 H_1(p)=-\sum_{i=1}^N p_i\log p_i.
\end{equation}
The base of the logarithm is arbitrary. In the following, we will denote $\log_D$ the base $D$ logarithm. The Tsallis entropy
is a simple transformation of the Rényi entropy, but is nonextensive. Often associated to these entropies, and central in the formulation of nonextensive statistical mechanics is the concept of escort distributions: if $\{p_i \}$ is the original distribution, then its escort distribution $P$ is defined by 
\begin{equation}
 P_i = \frac{p_i^q}{\sum_{i=1}^N p_i^q}.
\label{eq:escort}
\end{equation}
The parameter $q$ behaves as a microscope for exploring different regions of the measure $p$ \cite{chhabra_direct_1989}: for $q>1$, the more singular regions are amplified, while for $q<1$ the less singular regions are accentuated. The escort distributions have been introduced as a tool in the context of multifractals. Interesting connections with the standard thermodynamic are in \cite{chhabra_direct_1989,beck_thermodynamics_1993}. Discussion of their geometric properties can also be found in \cite{abe_geometry_2003}. It is also interesting to note that the escort distributions can be found as the result of a maximum entropy problem with a constraint on the expected value of a logarithmic quantity, see \cite[p. 53]{harte_multifractals:_2001} in the context of multifractals, or \cite{bercher_tsallis_2008} for a different view.  We shall also point out that the `deformed' information measure like the Rényi entropy (\ref{eq:renyi}) and the escort distribution (\ref{eq:escort}) are originally two distincts concepts, as indicated here by the different notations $\alpha$ and $q$. There is a lengthy discussion on this point in \cite{pennini_semiclassical_2007}. \\

In the information theory of communication, the entropy is the measure of the quantity of information in a message, and a primary aim is to represent the possible messages in an efficient manner, that is to find a compact representation of the information according to a measure of `compactness'. This is the role of source coding. 
In this note, we discuss the interest of escort distributions and alternative entropies in this context. This suggests possible connections between coding theory and the measure of complexity in nonextensive statistical mechanics. Related works are the study of generalized channel capacities \cite{landsberg_distributions_1998}, the notion of nonadditive information content \cite{yamano_information_2001}, the presentation of a generalized rate distorsion theory \cite{venkatesan_generalized_2009}.
The first section is devoted to a very short presentation of the source coding context, and to the presentation of the fundamental Shannon source coding theorem. In section \ref{sec:Campbell}, we describe a source coding theorem relating a new measure of length and the Rényi entropy. In the next section, we show that it is possible to obtain the very same optimum codes, as well as a practical procedure, using a reasoning based on the nonextensive generalized mean as the measure of length. In section  \ref{sec:yetanother}, we introduce another measure of length, involving escort distribution, and obtain general inequalities for this measure, where the lower bound, once again is a Rényi entropy. We show that the corresponding optimum codes are the standard Shannon codes. Finally, in section \ref{sec:connections} we discuss the connections between these different results.

\section{Source coding}
In source coding, one considers a set of symbols  $\mathcal{X}=\{x_1, x_2, \ldots x_N\}$, and a source that produces symbols $x_i$ from $\mathcal{X}$ with probabilities $p_i$ where $\sum_{i=1}^N p_i = 1$. The aim of source coding is to encode the source using an alphabet of size $D$, that is to map each symbol $x_i$ to a codeword $c_i$ of length $l_i$ expressed using the $D$ letters of the alphabet. It is known that if the set of lengths $l_i$ satisfies the Kraft-Mac Millan inequality 
\begin{equation}
 \sum_{i=1}^N D^{-l_i} \leq 1,
 \label{eq:kraft}
\end{equation}
then there exists a uniquely decodable code with these lengths, which means that any sequence $c_{i1}c_{i2}\ldots c_{in}$ can be decoded unambiguously into a sequence of symbols $x_{i1}x_{i2}\ldots x_{in}$. Furthermore, any uniquely decodable code satisfies the Kraft-Mac Millan inequality (\ref{eq:kraft}). The Shannon source coding theorem (noiseless coding theorem) indicates that the expected length of the code $\bar{L}$ is bounded below by the entropy of the source, $H_1(p)$, and that the best uniquely decodable code satisfies
\begin{equation}
 {H_1(p)} \leq \bar{L}=\sum_i p_i l_i < {H_1(p)} + 1,
\label{eq:shannon_theo}
\end{equation}
where the logarithm in the definition of the Shannon entropy is taken in base $D$. 
This result indicates that the Shannon entropy $H_1(p)$ is the fundamental limit on the minimum average length for any code constructed for the source. The lengths of the individual codewords, also called `bit-numbers' \cite[p. 46]{beck_thermodynamics_1993}, are given by 
\begin{equation}
 l_i = -\log_D p_i
\label{eq:bitnumberp}
\end{equation}
where $\log_D$ denotes the logarithm in base $D$. Obviously these code lengths enable to attain the entropy in the left of the inequality (\ref{eq:shannon_theo}). The characteristic of these optimum codes is that they assign the shorter codewords to the most likely symbols and the longer codewords to unlikely symbols.  The uniquely decodable code can be chosen to have the prefix property, i.e.
the property that no codeword is a prefix of another codeword.

\section{Source coding with Campbell measure of length}
\label{sec:Campbell}

It is well-known that Huffman coding yields a prefix code which minimizes the expected length and approaches the optimum limit $l_i=-\log_D p_i$. What is much less well known is that some other forms of lengths have been considered \cite{baer_source_2006}, the first and definitely fundamental contribution being the paper of Campbell \cite{campbell_coding_1965}. Since the codewords lengths obey to the relation (\ref{eq:bitnumberp}), low probabilities  yield very long words. But the cost of using a word is not necessarily a linear function of its length, and it is possible that adding a letter to a long word cost much more than adding a letter to a shorter word. This led Campbell to the proposal of a new average length measure, featuring an exponential account of the  elementary lengths of the codewords. This length, which is called a $\beta$-exponential mean or Campbell length, is a Kolmogorov-Nagumo generalized mean associated to an exponential function. It is defined by
\begin{equation}
 C_\beta=\frac{1}{\beta} \log_D \sum_{i=1}^N p_i D^{\beta l_i},
\label{eq:CambellLength}
\end{equation}
where $\beta$ is a strictly positive parameter. 
The remarkable result \cite{campbell_coding_1965} is that just as Shannon entropy is the lower bound on the average codeword length of an uniquely decodable code, the Rényi entropy of order $q$, with $q=1/(\beta+1)$, is the lower bound on the exponentially weighted codeword length (\ref{eq:CambellLength}):
\begin{equation}
 C_\beta \geq H_q(p).
\label{eq:Campbell_theo}
\end{equation}
A simple proof of this result will be given below. It is easy to check that the equality is achieved by choosing the $l_i$ such that
\begin{equation}
 D^{- l_i} = P_i = \frac{p_i^q}{\sum_{j=1}^N p_j^q},
\end{equation}
that is 
\begin{equation}
 l_i= -q \log_D p_i + (1-q) H_q(p).
\label{eq:opt_li_Pi}
\end{equation}
Obviously, the individual lengths obtained this way can be made smaller than the Shannon lengths $l_i=-\log_D p_i$, especially for small $p_i$, by selecting a sufficiently small value of $q$. Hence, the procedure effectively penalizes the longer codewords and yields a code different from Shannon's code, with possibly shorter codewords associated to the low probabilities. 

\section{Source coding with nonextensive generalized mean}
\label{sec:generalized}

In the standard measure of average length $\bar{L}=\sum_i p_i l_i$, we have a linear combination of the individual lengths, with the probabilities $p_i$ as weights. In order to increase the impact of the longer lengths with low probabilities, the Campbell's length uses an exponential of the length.  A different approach to the problem can be to modify the weigths in the linear combination, so as to raise the importance of the terms with low probabilities. A simple way to achieve this is to deform, flatten, the original probability distribution 
 and use the new distribution as weights rather than the $p_i$. Of course, a very good candidate is the escort distribution, which leads us to the `average length measure'
\begin{equation}
 M_q = \sum_{i=1}^N \frac{p_i^q}{\sum_j p_j^q} l_i = \sum_{i=1}^N P_i l_i,
\end{equation}
which is nothing but the generalized expected value of nonextensive statistical mechanics according to the third mean values' choice of Tsallis, Mendes and Plastino \cite{tsallis_role_1998}.   
For the virtual source with distribution $P$, the standard expected length is $M_q$, and the classical Shannon noiseless source coding theorem immediately applies, leading to
\begin{equation}
 M_q \geq H_1(P),
\label{eq:MqH1}
\end{equation}
with equality if
\begin{equation}
 l_i=-\log_D P_i
\label{eq:bitnumberP}
\end{equation}
which is exactly the lengths in (\ref{eq:opt_li_Pi}) obtained via Campbell's measure. This easy result has also be mentioned in \cite{yamano_information_2001}.\footnote{In this interesting paper, another inequality is given for the generalized mean: $M_q \geq S_q(p)$,
where $S_q$ is the normalized version of Tsallis entropy. In fact, this is only true under the condition $\sum_i \exp_q (-l_i) \leq 1$, with the equality occuring for $l_i=-\ln_q(p_i)$, where $\exp_q$ and $\ln_q$ denote the standard nonextensive $q$-deformed exponential and logarithm. When these lengths $l_i$ also fullfill the Kraft-Mac Millan inequality we have $M_q=S_q(p)>H_1(P)$.}

The simple relation $l_i=-\log_D P_i$ for the minimization of $M_q$ subject to the Kraft-Mac Millan inequality has a direct practical implication. Indeed, it suffices to feed a standard coding algorithm, namely a Huffman coder, with the escort distribution $P$ instead of the natural distribution $p$, to obtain as a result a code tailored for the Campbell's length measure $C_\beta$ or equivalently for the length measure $M_q$. 
 A simple example, with $D=2$, is reported in Table~\ref{tab:ComparWords}: we used a standard Huffman algorithm with the original distribution and the escort distributions with $q=0.7$ and $q=0.4$.  
\begin{center}
\begin{table*}[bht]
	\centering
\begin{tabular}{|l|l|l|l|} 
\hline
$p_i$ & $q=1$~~~ & $q=0.7$~~ & $q=0.4$~ \\ \hline
0.48 & 0 & 0 & 00 \\ 
0.3 & 10 & 10 & 01 \\ 
0.1 & 110 & 1100 & 100 \\ 
0.05 & 1110 & 1101 & 101 \\ 
0.05 & 11110 & 1110 & 110 \\ 
0.01 & 111110 & 11110 & 1110 \\ 
0.01 & 111111 & 11111 & 1111 \\  \hline
\end{tabular}
	\caption{\label{tab:ComparWords} Examples of codes in the binary case, for different values of $q$.}
\end{table*}
\end{center}

It is worth noting that some specific algorithms have been developed for Campbell's length \cite{humblet_generalization_1981,blumer_renyi_1988,baer_source_2006}. The remark above gives an easy alternative. An important point is that these new codes have direct applications: they are optimum for minimizing the probability of buffer overflows \cite{humblet_generalization_1981}, or, with $q>1$ for maximizing the chance of the reception of a message in a single snapshot \cite{baer_optimal_2008}. In the second case, the choice $q>1$ increases the main features of the probability distribution, then leading to select more short codewords for the highest probabilities; this maximizes the chance of a complete reception of a message in a single transmission of limited size.

\section{Another measure of length with Rényi bounds}
\label{sec:yetanother}

Given these results, it is now interesting to introduce a new measure of average length, similar to Campbell's length but mixing both a an exponential weight of individual lengths $l_i$ and an escort distribution. This measure is defined by
\begin{equation}
 L_q = \frac{1}{q-1} \log_D \left[ \sum_{i=1}^N \frac{p_i^q}{\sum_j p_j^q} D^{(q-1)l_i} \right].
\label{eq:Blength}
\end{equation}
Some specific values are as follows. It is easy to see that $L_0 = -\log_D \sum_i D^{-l_i} + \log_D N$. When $q \rightarrow +\infty$, the maximum of the probabilities, say $p_k=\mathrm{arg~max}_i p_i$ emerges, and $L_\infty = l_k$, where $l_k$ is the length associated to $p_k$, the maximum among the probabilities $p_i$. By L'Hospital's rule, we also obtain that $L_1=\bar{L}=\sum_i p_i l_i$.  As for Campbell's measure, it is possible to show that $L_q$ is bounded below by the Rényi entropy.

As in Campbell's original proof, let us consider the Hölder inequality 
\begin{equation}
         \biggl( \sum_{i=1}^N |x_i|^p \biggr)^{\!1/p\;} \biggl( \sum_{i=1}^N |y_i|^{p'} \biggr)^{\!1/{p'}} \le \sum_{i=1}^N |x_i\,y_i| \text{ for all sequences }(x_1,\ldots,x_N),(y_1,\ldots.y_N)\in\mathbb{R}^N
\end{equation}
for $p$ or ${p'}$ in $(0,1)$  and such that $1/p+1/{p'}=1$. Note that the reverse inequality is true when $p$ and ${p'}$ are in $[1,+\infty)$.
Suppose that the $l_i$ are the lengths of the codewords in a uniquely decodable code, which means that they satisfy the Kraft inequality (\ref{eq:kraft}). If we let now $x_i=p_i^\alpha D^{- l_i}$ and $y_i=p_i^{-\alpha}$, it comes
\begin{equation}
   \biggl( \sum_{i=1}^N p_i^{\alpha p} D^{-p l_i} \biggr)^{\!1/p\;} \biggl( \sum_{i=1}^N p_i^{-\alpha {p'}}  \biggr)^{\!1/{p'}} \le \sum_{i=1}^N D^{- l_i} \le 1,
\label{eq:ineqH}
\end{equation}
where the last inequality in the right is the Kraft inequality. 

If we let $\alpha p=1$, then $\alpha=-1/\beta$, and  $-\alpha {p'}=\alpha/(\alpha-1)=1/(\beta+1)$. Then, (\ref{eq:ineqH}) reduces to
\begin{equation}
   \biggl( \sum_{i=1}^N p_i D^{\beta l_i} \biggr)^{\!-1/\beta\;} \biggl( \sum_{i=1}^N p_i^{1/(\beta+1)}  \biggr)^{\!(\beta+1)/\beta} \le 1.
\end{equation}
Taking the base $D$ logarithm, we obtain the Campbell theorem $C_\beta \geq H_q(p)$, with $q=1/(\beta+1)$.

If we now take $\alpha p = q$ and choose $-\alpha {p'}=1$, we obtain
\begin{equation}
   \biggl( \sum_{i=1}^N p_i^{q} D^{-p l_i} \biggr)^{\!1/p\;}  \le 1,
\end{equation}
where we used of course the fact that the probabilities sum to one. The condition $1/p+1/p'=1$ easily gives $p=1-q$. Dividing the two sides by $(\sum_i p_i^q)^{1/(1-q)}$, taking the logarithm and changing the sign of the inequality, we finally obtain
\begin{equation}
 \frac{1}{q-1} \log_D \biggl( \sum_{i=1}^N \frac{p_i^q}{\sum_j p_j^q} D^{(q-1) l_i} \biggr)  \geq  \frac{1}{1-q} \log_D \sum_{i=1}^N p_i^q 
\end{equation}
which gives the simple inequality
\begin{equation}
 L_q \geq H_q.
\label{eq:Btheorem}
\end{equation}
Hence we obtain that the new length measure of order $q$ is lower bounded by the Rényi entropy of the same order. Note that this result include Shannon result in the special case $q=1$. Interestingly, it is easy to check that we have equality in (\ref{eq:Btheorem}) for $l_i=-\log_D p_i$, which is nothing but the optimal lengths in the Shannon coding theorem. Hence, it is remarkable that the whole series of inequalities (\ref{eq:Btheorem}) become equalities for the choice $l_i=-\log_D p_i$ which appears as a kind of universal value in this context. 

This result can draw attention to alternative coding algorithms, based on the minimization of $L_q$, or alternative characterizations of the optimal code. For instance, the inequality (\ref{eq:Btheorem})  shows, as a direct consequence, that the Shannon code with $l_i=-\log_D p_i$ minimizes the length of the codeword associated to the maximum probability. Indeed, when $q\rightarrow+\infty$, $L_\infty \rightarrow l_k$ the length of the codeword of maximum probability, and  $L_\infty$ is minimum when $l_k$ has its minimum value $H_\infty=-\log_D p_k$.

Since the Rényi and Tsallis entropy are related by a simple monotone transformation, inequalities similar to (\ref{eq:Campbell_theo}) and (\ref{eq:Btheorem}) exist with Tsallis entropy bounds. 

\section{Connections between the different length measures}
\label{sec:connections}

It is finally useful to exhibit an interplay between the two length measures, their minimizers, and the standard and escort distributions. The Campbell measure in (\ref{eq:CambellLength}) involves the distribution $p$, an exponential weight with index $\beta$. The optimum lengths that achieve the equality in the inequality (\ref{eq:Campbell_theo}) are the bit-numbers associated to the escort distribution $l_i=-\log_D P_i$. On the other hand, the measure (\ref{eq:Blength}) involves the escort distribution $P$ instead of $p$, has an index $q$ and the optimum lengths that achieve the equality in the extended source coding inequality (\ref{eq:Btheorem}) are the bit-numbers $l_i=-\log_D p_i$ associated to the original distribution. We know that the transformation $q \leftrightarrow 1/q$ \cite[p. 543]{tsallis_role_1998} links the original and escort distribution, that is the distribution $p$ is the escort distribution with index $1/q$ of the distribution $P$. This remark enables to find an equivalence between thermostatistics formalisms base on linear and generalized averages \cite{raggio_equivalence_1999,naudts_dual_2002}. Here, when we substitute $q$ by $1/q$ in (\ref{eq:Blength}), and therefore $P$ by $p$, we end with Campbell length (\ref{eq:CambellLength}) where $q=1/(\beta+1)$. Concerning the entropy bound in (\ref{eq:Campbell_theo}) and (\ref{eq:Btheorem}), we shall also observe that $H_{\frac{1}{q}}(P)=H_q(p)$, so that we have finally equivalence between the two inequalities (\ref{eq:Campbell_theo}) and (\ref{eq:Btheorem}). This is a new illustration of the duality between standard and escort distributions. 

As a last remark, let us mention that if we apply Jensen inequality to the exponential function in the sum defining $L_q$ (\ref{eq:Blength}), we then obtain $M_q \geq L_q$, where $M_q$ is the generalized mean, taken with respect to the escort distribution, and we have
\begin{equation}
 M_q \geq L_q \geq H_q.
\end{equation}
The equality in $M_q \geq L_q$ means that the transformation in Jensen inequality is a straight line, which means $q=1$. In such case, we still obtain $M_1 \geq H_1(p)$, which is nothing but the standard Shannon coding theorem.

\section{Conclusions}

In this Letter, we have pointed out the relevance of Rényi entropy and escort distributions in the context of source coding. This suggests  possible connections between coding theory and the main tools of nonextensive statistical mechanics. We have first outlined an overlooked result by Campbell that gave the first operational characterization of Rényi entropy, as the lower bound in the minimization of a deformed measure of length. We then considered some alternative definitions of measure of length. 
We showed that Campbell's optimum codes can also be obtained using another natural measure of length based on escort distributions. Interestingly, this provides an easy practical procedure for the computation of these codes. Next, we introduced a third measure of length involving both an exponentiation, as in Campbell's case, and escort distributions. We showed that this length is also bounded below by a Rényi entropy. Finally, we showed that the duality between standard and escort distributions connects some of these results. 

Further work should consider the extension of these results, namely the new lengths definitions, in the context of channel coding. With these new lengths, we also intend to investigate the problem of model selection, as in Rissanen MDL (Minimum Description Length) procedures.


%

\begin{thebibliography}{10}
\expandafter\ifx\csname url\endcsname\relax
  \def\url#1{\texttt{#1}}\fi
\expandafter\ifx\csname urlprefix\endcsname\relax\def\urlprefix{URL }\fi

\bibitem{tsallis_introduction_2009}
C.~Tsallis, Introduction to {N}onextensive {S}tatistical {M}echanics, 1st
  Edition, Springer, 2009.

\bibitem{harte_multifractals:_2001}
D.~Harte, Multifractals: Theory and Applications, 1st Edition, Chapman \&
  {Hall}, 2001.

\bibitem{jizba_world_2004}
P.~Jizba, T.~Arimitsu, Ann Phys 312 (2004) 17.

\bibitem{chhabra_direct_1989}
A.~Chhabra, R.~V. Jensen, Phys Rev Lett 62 (1989) 1327.

\bibitem{beck_thermodynamics_1993}
C.~Beck, F.~Schloegl, Thermodynamics of Chaotic Systems, Cambridge University
  Press, 1993.

\bibitem{abe_geometry_2003}
S.~Abe, Phys Rev E 68 (2003) 031101.

\bibitem{bercher_tsallis_2008}
J.-F. Bercher, Phys Lett A 372 (2008) 5657.

\bibitem{pennini_semiclassical_2007}
F.~Pennini, A.~Plastino, G.~Ferri, Physica A: Statistical Mechanics and its
  Applications 383 (2007) 782.

\bibitem{landsberg_distributions_1998}
P.~T. Landsberg, V.~Vedral, Phys Lett A 247 (1998) 211.

\bibitem{yamano_information_2001}
T.~Yamano, Phys Rev E 63 (2001) 046105.

\bibitem{venkatesan_generalized_2009}
R.~Venkatesan, A.~Plastino, Physica A 388 (2009) 2337.

\bibitem{baer_source_2006}
M.~Baer, IEEE Trans Inf Theory 52 (2006) 4380.

\bibitem{campbell_coding_1965}
L.~L. Campbell, Inf Control 8 (1965) 423.

\bibitem{tsallis_role_1998}
C.~Tsallis, R.~S. Mendes, A.~R. Plastino, Physica A 261 (1998) 534.

\bibitem{humblet_generalization_1981}
P.~Humblet, IEEE Trans Inf Theory 27 (1981) 230.

\bibitem{blumer_renyi_1988}
A.~Blumer, R.~{McEliece}, IEEE Trans Inf Theory 34 (1988) 1242.

\bibitem{baer_optimal_2008}
M.~Baer, IEEE Trans Inf Theory 54 (2008) 1273.

\bibitem{raggio_equivalence_1999}
G.~A. Raggio, On equivalence of thermostatistical formalisms,
  http://arxiv.org/abs/cond-mat/9909161 (1999).

\bibitem{naudts_dual_2002}
J.~Naudts, Chaos Solitons Fractals 13 (2002) 445.

\end{thebibliography}

%
%
%
%

 \end{document}